# Controlling structure and interfacial interaction of monolayer TaSe$_2$ on bilayer graphene

Hyobeom Lee[1†], Hayoon Im[2†], Byoung Ki Choi[3], Kyoungree Park[1], Yi Chen[4,5,6,7], Wei Ruan[4,8], Yong Zhong[9,10], Ji-Eun Lee[3,11], Hyejin Ryu[12], Michael F. Crommie[4,13], Zhi-Xun Shen[9,10], Choongyu Hwang[2*], Sung-Kwan Mo[3*], and Jinwoong Hwang[1*]

[1]Department of Physics and Institute of Quantum Convergence Technology, Kangwon National University, Chuncheon, South Korea

[2]Department of Physics, Pusan National University, Busan, South Korea

[3]Advanced Light Source, Lawrence Berkeley National Laboratory, Berkeley, CA, USA

[4]Department of Physics, University of California, Berkeley, CA, USA

[5]International Center for Quantum Materials, School of Physics, Peking University, Beijing 100871, China

[6]Collaborative Innovation Center of Quantum Matter, Beijing 100871, China

[7]Interdisciplinary Institute of Light-Element Quantum Materials and Research Center for Light-Element Advanced Materials, Peking University, Beijing 100871, China

[8]State Key Laboratory of Surface Physics, New Cornerstone Science Laboratory, and Department of Physics, Fudan University, Shanghai, China

[9]Geballe Laboratory for Advanced Materials, Department of Physics and Applied Physics, Stanford University, Stanford, CA, USA

[10]Stanford Institute for Materials and Energy Sciences, SLAC National Accelerator Laboratory, Menlo Park, CA, USA

[11]Max Planck POSTECH Center for Complex Phase Materials, Pohang University of Science and Technology, Pohang, South Korea

[12]Center for Spintronics, Korea Institute of Science and Technology, Seoul, South Korea

[13]Materials Sciences Division, Lawrence Berkeley National Laboratory, Berkeley, CA, USA

† Hyobeom Lee and Hayoon Im contributed equally to this work.

* Corresponding authors: jwhwang@kangwon.ac.kr, skmo@lbl.gov, ckhwang@pusan.ac.kr

**Abstract**

Tunability of interfacial effects between two-dimensional (2D) crystals is crucial not only for understanding the intrinsic properties of each system, but also for designing electronic devices based on ultra-thin heterostructures. A prerequisite of such heterostructure engineering is the availability of 2D crystals with different degrees of interfacial interactions. In this work, we report a controlled epitaxial growth of monolayer $TaSe_2$ with different structural phases, 1$H$ and 1$T$, on a bilayer graphene (BLG) substrate using molecular beam epitaxy, and its impact on the electronic properties of the heterostructures using angle-resolved photoemission spectroscopy. 1$H$-$TaSe_2$ exhibits significant charge transfer and band hybridization at the interface, whereas 1$T$-$TaSe_2$ shows weak interactions with the substrate. The distinct interfacial interactions are attributed to the dual effects from the differences of the work functions as well as the relative interlayer distance between $TaSe_2$ films and BLG substrate. The method demonstrated here provides a viable route towards interface engineering in a variety of transition-metal dichalcogenides that can be applied to future nano-devices with designed electronic properties.

**Introduction**

The exotic properties of atomically thin two-dimensional (2D) crystals, first revealed in graphene, have led to a tremendous expansion in the 2D materials research [1-4]. In particular, controllable atomic layer-by-layer growth using chemical vapor deposition and molecular beam epitaxy (MBE) has allowed us to address fundamental issues in the 2D limit and to search for artificial interfaces with designed functionalities [2-8]. Transition-metal dichalcogenides (TMDCs) provide a fertile platform to realize a number of exotic properties with various constituent atoms and crystal structures [1-3], e.g., $1H$ (trigonal prismatic coordination) and $1T$ (octahedral coordination) with differences in the coordination of six chalcogen atoms surrounding a metal atom. One caveat, and simultaneously an advantage of 2D crystals, is that the intrinsic physical properties of epitaxially grown monolayer (ML) TMDC films can be modified by strong interactions with a substrate [9-13]. Bilayer graphene (BLG) on a SiC(0001) substrate has been ubiquitously used for the epitaxial growth of layered 2D materials when studying the intrinsic characteristics of van der Waals (vdW) materials in a 2D limit due to relative chemical inertness of BLG [14-18]. The weak interactions between BLG and epitaxial vdW materials can preserve the intrinsic properties of overlaid 2D materials. Indeed, the formation of novel ground states has been demonstrated in TMDCs by using BLG substrate, e.g., the indirect-to-direct band gap transition in $2H$-MoSe$_2$ [14], the exciton condensed states in ML $1T$-ZrTe$_2$ [15], the quantum spin Hall state in ML $1T$'-WTe$_2$ [16], and metal-to-insulator transition in $1T$-IrTe$_2$ [17].

Among the family of TMDCs, MBE-grown MX$_2$ (M = Nb, Ta; X = S, Se) on a BLG substrate has been intensively studied, and the growth recipes have been well established [19], making them a great platform to study exotic quantum phenomena in the ML regime. Examples include charge density waves (CDW) and Ising superconductivity in $1H$-MX$_2$ [20-22], exotic orbital textures with Mott insulating states and quantum spin liquid behavior in $1T$-MX$_2$ [23-25], and heavy fermionic behaviors in $1T/1H$-MX$_2$ heterostructures [24-28]. One critical aspect to consider but often neglected is that BLG substrate may give a significant charge transfer to the overlaid MX$_2$ films due to a substantial difference in work functions between MX$_2$ and BLG, which may strongly affect the intrinsic properties of ML MX$_2$ [29-31]. Considering that the ground states of atomically thin TMDC films can be easily modified by the amount of extra charge doping [11,15,32], it is crucial to carefully study the effect of the BLG substrate on overlaid ML MX$_2$ films.

Here, we report the electronic structure of epitaxially grown ML TaSe$_2$ films on a BLG substrate using angle-resolved photoemission spectroscopy (ARPES). The interfacial interactions have been modified through the selective growth of structural phases ($1T$ and $1H$) of ML TaSe$_2$ using MBE. Strong interactions between ML $1H$-TaSe$_2$ and BLG were evidenced by kinked band structures and significant charge transfer from BLG to TaSe$_2$, while weakly interacting ML $1T$-TaSe$_2$ on BLG does not exhibit any charge transfer or band hybridization. The former deviates from the previous works that found the quasi-freestanding nature of MBE-grown ML TMDC on BLG [14-18]. Scanning tunneling microscopy (STM) measurements and first-principles calculations reveal differences in the atomic height and the modified work functions in the ML limit of two phases of TaSe$_2$, resulting in different electronic responses at the interface.

**Results**

Figure 1**a** presents the schematics for the controlled growth of ML TaSe$_2$ on a BLG substrate using MBE. It is well known that $1H$- and $1T$-TaSe$_2$ films can be selectively

synthesized on BLG by controlling substrate temperature ($T_{growth}$) during the growth; low and high $T_{growth}$ are suitable for the formation of 1$H$-TaSe$_2$ and 1$T$-TaSe$_2$, respectively [19]. Figures 1**b** and 1**d** show the ARPES spectra of MBE-grown ML TaSe$_2$ depending on $T_{growth}$. ARPES intensity maps demonstrate that the ML TaSe$_2$ film grown at high $T_{growth}$ (= 750 ˚C) shows an insulating band structure (Fig. 1**b**) while the low $T_{growth}$ (= 450 ˚C) shows metallic behavior (Fig. 1**d**). These results are consistent with the Mott insulating state by the Star-of-David (SoD) CDW transition in ML 1$T$-TaSe$_2$ and the metallic nature of ML 1$H$-TaSe$_2$, respectively [19,22,23]. On the other hand, the ML TaSe$_2$ film grown at intermediate $T_{growth}$ (= 600 ˚C) exhibits mixed band structures of ML 1$H$- and 1$T$-TaSe$_2$ (Fig. 1**c**).

The selective fabrication of ML TaSe$_2$ films by controlling $T_{growth}$ is also confirmed by core-level measurements since the change of crystal structures generates different crystal fields in TaSe$_2$ [23,33,34]. Figures 1**e** and 1**f** represent core-level spectra for Ta 4$f$ and Se 3$d$, respectively. The peak shapes and positions of Ta 4$f$ and Se 3$d$ obtained from high $T_{growth}$ = 750 ˚C (light blue) and low $T_{growth}$ = 450 ˚C (dark blue) are in agreement with ones of 1$T$- and 1$H$-TaSe$_2$, respectively, as reported [35,36]. On the other hand, for moderate $T_{growth}$ = 600 ˚C, not only do multiple peaks appear in both Ta 4$f$ and Se 3$d$, but they also have the same positions with the core peaks from 1$T$- and 1$H$-TaSe$_2$, indicating the coexistence of the 1$H$- and 1$T$-TaSe$_2$ islands. ARPES and core-level measurements demonstrate the importance of delicate control of $T_{growth}$ to tune the structural phases of ML TaSe$_2$ on a BLG substrate [19,23].

To investigate the effect of the BLG substrate on ML TaSe$_2$, the BLG π bands have been measured with and without overlaid TaSe$_2$ [37-40]. Figure 2**a** shows an ARPES intensity map of the BLG π bands without TaSe$_2$ taken at the $K_G$ point perpendicular to the $\Gamma_G$–$K_G$ direction of the Brillouin zone (BZ) of BLG. The obtained as-grown BLG π bands are intrinsically doped by electrons due to the presence of the SiC substrate [41]. The Dirac energy ($E_D$), defined here as the middle of the conduction band minimum and the valence band maximum, is located at ~0.3 eV below Fermi energy ($E_F$) extracted from the 2$^{nd}$ derivative ARPES spectrum (red lines) as shown in Fig. 2**d**. Figures 2**b** and 2**e** present the BLG π bands taken from fully covered ML 1$T$-TaSe$_2$ films. Compared to as-grown BLG on an SiC substrate (Fig. 2**a**), there are two non-dispersive states with weak spectral intensity located at 0.3 eV and 0.9 eV below $E_F$, which originate from ML 1$T$-TaSe$_2$ due to SoD CDW transition [42]. Although the additional bands are crossing the BLG π bands, the BLG π band dispersion is hardly changed. Moreover, we found a small amount of charge transfer from BLG to ML 1$T$-TaSe$_2$, i.e., a slight shift of $E_D$ from 0.30 eV to 0.24 eV below $E_F$ (Fig. 2**e**), indicating weak interactions between ML 1$T$-TaSe$_2$ and BLG.

On the other hand, remarkable changes are observed in BLG π bands when ML 1$H$-TaSe$_2$ is grown on a BLG substrate. As shown in Figs. 2**c** and 2**f**, ARPES intensity maps do not show the valence band maximum and $E_D$ of BLG π bands. Extended straight lines over the upper π band give $E_D$ at 0.135 eV above $E_F$. This result provides direct evidence of significant charge transfer from BLG to overlaid ML 1$H$-TaSe$_2$ [38,39]. Moreover, BLG π bands show kinked structures at the crossing points with Ta 5$d$ bands of 1$H$-TaSe$_2$ located at 0.1 eV and 0.38 eV below $E_F$ as denoted by orange and red dashed circles and arrows (Fig. 2**f**).

The charge transfer and the kinked structure are clearly resolved when the BLG π bands are taken along the $K_G$–$M_G$–$K_G$ direction of the BZ of BLG. Figure 3**a** shows ARPES intensity maps of BLG π bands for 0.5 ML 1$T$-TaSe$_2$ on a BLG substrate, i.e., 50% of partial coverage of the substrate by 1$T$-TaSe$_2$. The coverage of ML TaSe$_2$ films was determined by comparing reflection high-energy electron diffraction (RHEED) intensity ratio between BLG and TaSe$_2$ peaks. As obtained in Figs. 2**b** and 2**e**, the BLG π bands do not show any kinked structure at the crossing points with ML 1*T*-TaSe$_2$ bands, and there are just two branches of BLG π bands due to the presence of two layers of graphene [43]. We did not find any additional split of the BLG π band (Figs. 3**a** and 3**d**), indicating negligible interactions. On the other hand, the 0.5 ML 1*H*-TaSe$_2$ sample exhibits three branches of BLG π bands as denoted by yellow arrows in Figs. 3**b** and 3**e**. These multiple branches stem from the partial coverage (0.5 ML) of 1*H*-TaSe$_2$ films on BLG substrate and ARPES measurements simultaneously catch BLG π bands from both as-grown BLG/SiC(0001) and 1*H*-TaSe$_2$ on BLG/SiC(0001) due to finite spot size of the photon beam [18,23,33]. Indeed, for the nearly full coverage of 1*H*-TaSe$_2$ on a BLG substrate (Figs. 3**c** and 3**f**), the BLG π bands are reduced to two branches, which are shifted toward $E_F$ because of charge transfer from BLG to ML 1*H*-TaSe$_2$. Concomitantly, there is a discontinuity in the upper π band of BLG at 1.5 eV below $E_F$, as denoted by black-dashed circles in Figs. 3**b-c** and 3**e-f**. Such changes, e.g., charge transfer and kinked structures, indicate that there exist strong interactions between ML 1*H*-TaSe$_2$ and a BLG substrate [37-41].

BLG π bands at the $M_G$ point reveal another intriguing evidence of the charge transfer between ML 1*H*-TaSe$_2$ and a BLG substrate. We found that the split of upper and lower BLG π bands shows different split energy values (Δ$E$) depending on the overlaid ML TaSe$_2$ crystal structures. The split of the lower two branches in Fig. 3**e** has Δ$E$ = 0.38 eV, which is comparable to one of the as-grown BLG π bands on an SiC substrate [43] and of ML 1*T*-TaSe$_2$ on BLG (Fig. 3**d**). On the other hands, the upper two branches of BLG π bands in Fig. 3**e** has Δ$E$ = 0.50 eV, which corresponds to the hole-doped ML 1*H*-TaSe$_2$ on BLG (Fig. 3**f**). The enhanced Δ$E$ may originate from the inequivalent charge distribution in the upper and lower BLG layers [43,44]. While the lower graphene layer takes the electrons from the SiC substrate, the upper layer transfers the electrons to ML 1*H*-TaSe$_2$ [44], as evidenced in ARPES results (Figs. 2 and 3). The sufficient asymmetry of the charge density between the BLG layers induces the field at the respective interfaces, resulting in the enhancement of $\Delta E$ [44].

**Discussion**

The selective interactions in ML TaSe$_2$ films on BLG are non-trivial, because it is reasonable to expect similar amount of charge transfer in both structural phases of TaSe$_2$, considering the work function difference between BLG (4.3 eV) and bulk TaSe$_2$ (5.1 eV for 1$T$ and 5.5 eV for 2$H$) [45-47]. However, the work function can be modified when TaSe$_2$ is thinned down to ML [46-50]. The calculated work functions for 1$H$-TaSe$_2$ are hardly changed from bulk (5.5 eV) to ML (5.45 eV), whereas the work function of 1$T$-TaSe$_2$ are significantly reduced from bulk (5.10 eV) to ML (4.66 eV) (Fig. 4**a**). The difference in the charge transfer between TaSe$_2$ and BLG is due to the distinct behavior of the work function in the 2D limit of 1$T$ and 1$H$ phases of TaSe$_2$.

In addition, an interlayer distance between TaSe$_2$ and BLG can also play a crucial role in the electronic properties at the interface, since the Schottky barrier is modified as a function of the distance of vdW layers [51-55]. Our STM measurements reveal that MBE-grown ML 1$T$- and 1$H$-TaSe$_2$ on a BLG substrate show a different height of 1.02 nm and 0.85 nm, respectively (Figs. 4**b** and 4**c**). In general, height estimated from STM topography reflects atomic positions in real space as well as contributions from electronic structures. A height difference of 1.7 Å in STM data thus implies either that the vdW gap between ML 1$T$-TaSe$_2$ and BLG is wider by ~ 1.7 Å or that 1$H$-TaSe$_2$ has much lower density of states (DOS) so that the tip must move towards the 1$H$-TaSe$_2$ film (compared to 1$T$-TaSe$_2$) to maintain the same tunneling condition at certain sample bias voltage ($V_b$) [56]. Since the DOS taken at $V_b = -1$ V is larger in ML 1$H$-TaSe$_2$ than ML 1$T$-TaSe$_2$ [25,57,58], however, the obtained STM heights provide evidence of the shorter vdW gap between ML 1*H*-TaSe$_2$ and BLG, compared to that of ML 1*T*-TaSe$_2$. Hence, our findings suggest that the strong (weak) interactions between ML 1*H* (1*T*)-TaSe$_2$ and a BLG substrate originate from the dual effects of the significant (small) work function difference and the relatively shorter (larger) interlayer distances.

**Conclusions**

In summary, we have investigated the electronic structure of the ML TaSe$_2$ on BLG when the structural phase of TaSe$_2$ is selectively grown in a controlled way. The presence of ML 1*H*-TaSe$_2$ on BLG results in strong interactions evidenced by the energy shift due to hole doping in the BLG band structure and the kinked structure at the band crossing points between ML 1*H*-TaSe$_2$ and BLG. On the other hand, the presence of ML 1*T*-TaSe$_2$ on BLG shows nearly negligible effects on the BLG band structure, indicating weak interactions. The distinct response from ML 1*H*- and 1*T*- TaSe$_2$ on BLG originate from reduced interfacial distance and strongly reduced work function of 1*H*-TaSe$_2$ in the ML limit. Our findings provide an exceptional example of strong interactions between the BLG substrate and an epitaxially-grown TMDC material, which paves the way for discovering and manipulating novel electronic phases in 2D vdW materials and their heterostructures.

**List of abbreviations**

*2D*: two-dimension

**MBE**: molecular beam epitaxy

*TMDC*: transition metal dichalcogenides

*BLG*: bilayer graphene

*vdW*: van der Waals

*ML*: monolayer

*CDW*: charge density wave

*ARPES*: angle-resolved photoemission spectroscopy

*STM*: scanning tunneling microscopy

$T_{growth}$: substrate temperature during the growth

*SoD*: Star-of-David

*BZ*: Brillouin zone

$E_D$: Dirac energy

$E_F$: Fermi energy

$\Delta E$: split energy values

$V_b$: bias voltage

## Methods

### Thin film growth and in-situ ARPES measurement

The BLG substrate was prepared by flashing annealing of the 6H-SiC(0001) at 1300 °C for 60 cycles. The ML 1*H*- and 1*T*-TaSe$_2$ films were grown by molecular beam epitaxy (MBE) on epitaxial bilayer graphene on 6H-SiC(0001). The base pressure of the MBE chamber was 3 × 10$^{-10}$ Torr. High-purity Ta (99.99%) and Se (99.999%) were evaporated from an e-beam evaporator and a standard Knudsen effusion cell, respectively. The flux ratio was fixed as Ta:Se = 1:10, and the BLG substrate temperatures were ranged from 450 ˚C (1*H*-TaSe$_2$) to 750 ˚C (1*T*-TaSe$_2$). This yields a growth rate of 40 mins per ML monitored by *in situ* Reflection high-energy electron diffraction (RHEED).

The MBE-grown ML TaSe$_2$ films were transferred directly into the ARPES analysis chamber for the measurement at the HERS endstation of Beamline 10.0.1, Advanced Light Source, Lawrence Berkeley National Laboratory. ARPES data were taken using a Scienta R4000 analyzer at base pressure 3 × 10$^{-11}$ Torr. The photon energies were set at 50 eV for *s*-polarizations and 63 eV for *p*-polarizations with energy and angular resolution of 10–20 meV and 0.1°, respectively. The spot size of the photon beam on the sample was ~100 μm × 100 μm. Se capping layers of ~100 nm were deposited onto ML TaSe$_2$ films at room temperature to prevent contamination during transport through air to the ultrahigh vacuum (UHV) scanning tunneling microscopy (STM) chamber. Se capping layers were removed by annealing the sample to 200 ˚C overnight in the UHV before the STM measurements.

**STM measurement**

STM measurements are performed using a commercial Omicron LT-STM/AFM under UHV conditions at $T$ = 5 K with tungsten tips. STM topography was obtained in constant-current mode. STM tips were calibrated on an Au(111) surface by measuring the Au(111) Shockley surface state before all STS measurements. STS was performed under open feedback conditions by lock-in-detection of an alternating-current tunnel current with a small bias modulation at 401 Hz added to the tunneling bias. WSxM software was used to process the STM images.

**Density functional theory calculation**

Work function calculations were conducted using the density functional theory method with the Quantum ESPRESSO package [59]. We employed the generalized gradient approximation (GGA) of Perdew, Burke, and Ernzerhof (PBE) functionals [60]. A plane wave kinetic energy cutoff of 100 Ry (1360 eV) and 12 × 12 × 1 Monkhorst-Pack mesh were employed [61]. A vacuum gap thickness of 20 Å was introduced at the side of the slab for all systems to calculate the work function ($\phi = V_{vac} - E_F$). All work function values were extracted from the plane-averaged electrostatic potential.

**Data availability**

The data that support the plots within this paper and other findings of this study are available from the corresponding authors upon reasonable request.

**Acknowledgments**

The work at ALS is supported by the US DoE Office of Basic Energy Science under contract No. DE-AC02-05CH11231. The work at UC Berkeley (STM) is supported by the US National Science Foundation grant No. DMR-2221750. The work at KNU is supported by the National Research Foundation of Korea (NRF) grant funded by the Korean government (MSIT) (RS-2023-00280346), the GRDC (Global Research Development Center) Cooperative Hub Program through the NRF funded by the Ministry of Science and ICT (MIST) (RS-2023-00258359), and Semiconductor R&D Support Project through the Gangwon Technopark (GWTP) funded by Gangwon Province (No. GWTP 2023-027). The work at PNU is supported by the NRF of Korea (No. 2021R1A2C1004266 and No. RS-2023-00221154) and the National Research Facilities and Equipment Center (NFEC) grant funded by the Ministry of Education (No. 2021R1A6C101A429). Max Planck POSTECH/Korea Research Initiative is supported by the NRF of Korea (2022M3H4A1A04074153). J.-E.L. is partially supported by the ALS collaborative Postdoctoral Fellowship. Y.C. acknowledges support from the National Natural Science Foundation of China (Grant No. 12250001 and 92365114). W.R. acknowledges financial support from the National Science Foundation of China (Grant No. 12274087) and Shanghai Science and Technology Development Funds (Grant No. 22QA1400600). B.K.C. acknowledges supports from NRF (2021R1A6A3A14040322). The authors acknowledge the Urban Big data and AI Institute of the University of Seoul supercomputing resources. H.R. acknowledges support from KIST Institutional Program (2E32951) and NRF grant funded by the Korea government (MSIT) (No. 2021R1A2C2014179 and 2020R1A5A1016518)

**Author contributions**

JH, CH, SKM proposed and designed the research. JW performed film growth with assistance from HL, HI, KP, and L-EL. JW, HL, HI, YZ, and HR carried out the ARPES measurements and analyzed the ARPES data with assistance from CH, ZXS, and SKM. YC and WR carried out the STM measurements with assistance from MFC. BKC performed the DFT calculations. HL, HI, and JH wrote the manuscript and revised it with assistance from CH and SKM. All authors contributed to the scientific planning and discussions.

**Competing interests**

The authors declare no competing interests.

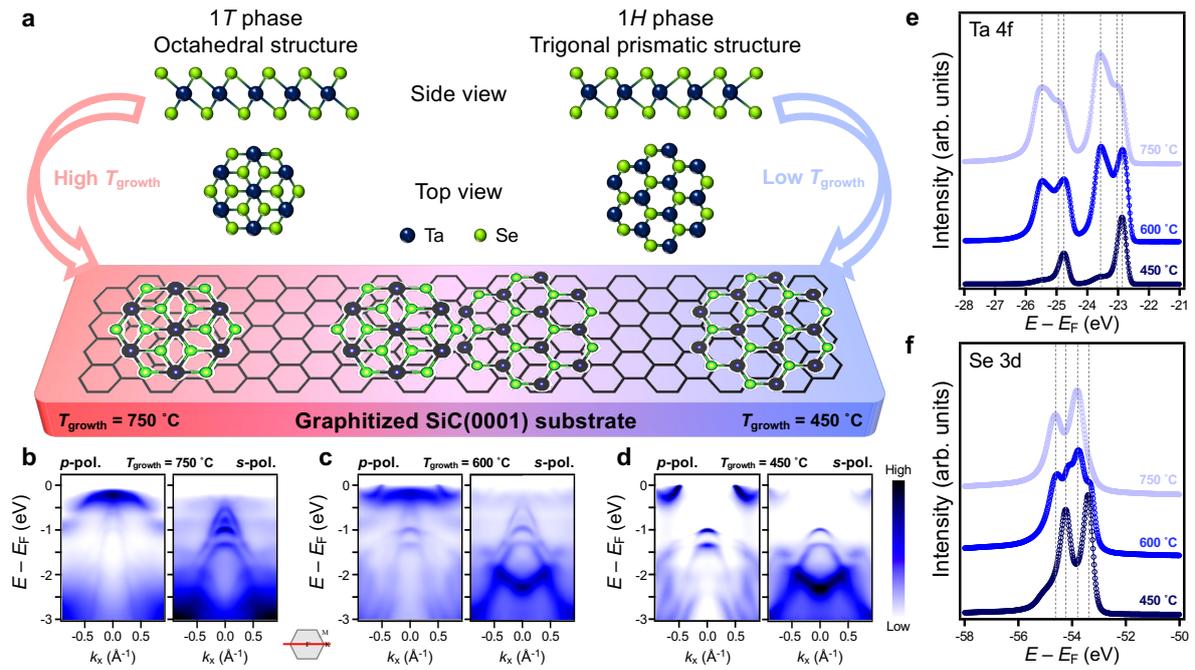

**Figure 1 Selective fabrications of 1*T*- and 1*H*-TaSe$_2$ on a BLG substrate**. **a**, Schematics of (top) Side- and top-view of atomic structures of TaSe$_2$ and (bottom) the $T_{\text{growth}}$ dependent TaSe$_2$ film synthesis on a BLG substrate. **b-d**, ARPES intensity maps of ML TaSe$_2$ films with three different $T_{\text{growth}}$. The formation of (**b**) ML 1*T*-TaSe$_2$ at $T_{\text{growth}}$ = 750 ˚C, (**c**) mixed structures of ML TaSe$_2$ at $T_{\text{growth}}$ = 600 ˚C, and (**d**) ML 1*H*-TaSe$_2$ at $T_{\text{growth}}$ = 450 ˚C. The *p*- and *s*-polarized ARPES intensity maps were taken with 63 eV and 50 eV photons, respectively, at 10 K. **e-f**, Core-level photoemission spectra from (**e**) Ta 4*f*- and (**f**) Se 3*d*-levels of ML TaSe$_2$ films. All the data were taken at 10 K.

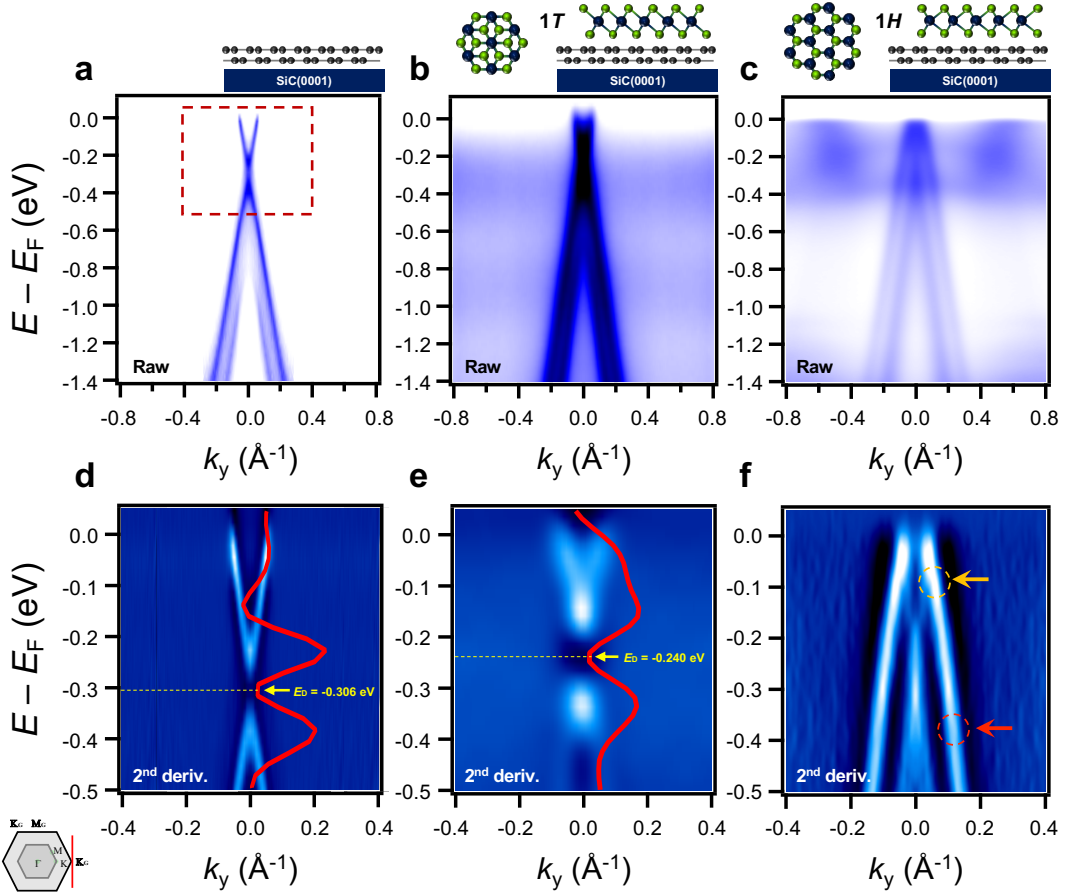

**Figure 2 ARPES spectra of BLG π bands with and without overlaid ML TaSe$_2$ films.** **a-c**, ARPES intensity maps of (**a**) as-grown BLG on SiC(0001), and BLG π bands covered with ML (**b**) 1$T$- and (**c**) 1$H$-TaSe$_2$ on a BLG substrate, respectively, taken at the K point of the BLG BZ (K$_G$) perpendicular to the Γ–K direction using $p$-polarized photons at 10 K. **d-f**, Second derivative of the zoomed-in ARPES intensity maps (dented by the red-dashed rectangle in panel **a**) for (**d**) as-grown BLG on SiC(0001), and BLG π bands covered with ML (**e**) 1$T$- and (**f**) 1$H$-TaSe$_2$ on a BLG substrate, respectively. Two non-dispersive bands with broad and weak spectral intensity at ~0.3 eV and ~0.9 eV below $E_F$ in **b** originate from ML 1$T$-TaSe$_2$. The red curves in panels **d** and **e** are energy distribution curves of the second derivative maps taken at $k_y = 0.0$ Å$^{-1}$. The yellow dashed lines and arrows indicate $E_D$. The orange and red dashed circles and arrows in panel **f** represent kinked structures of BLG π bands. M$_G$ and K$_G$ (M and K points of the BLG BZ) in the inset indicates the high symmetry points of BLG.

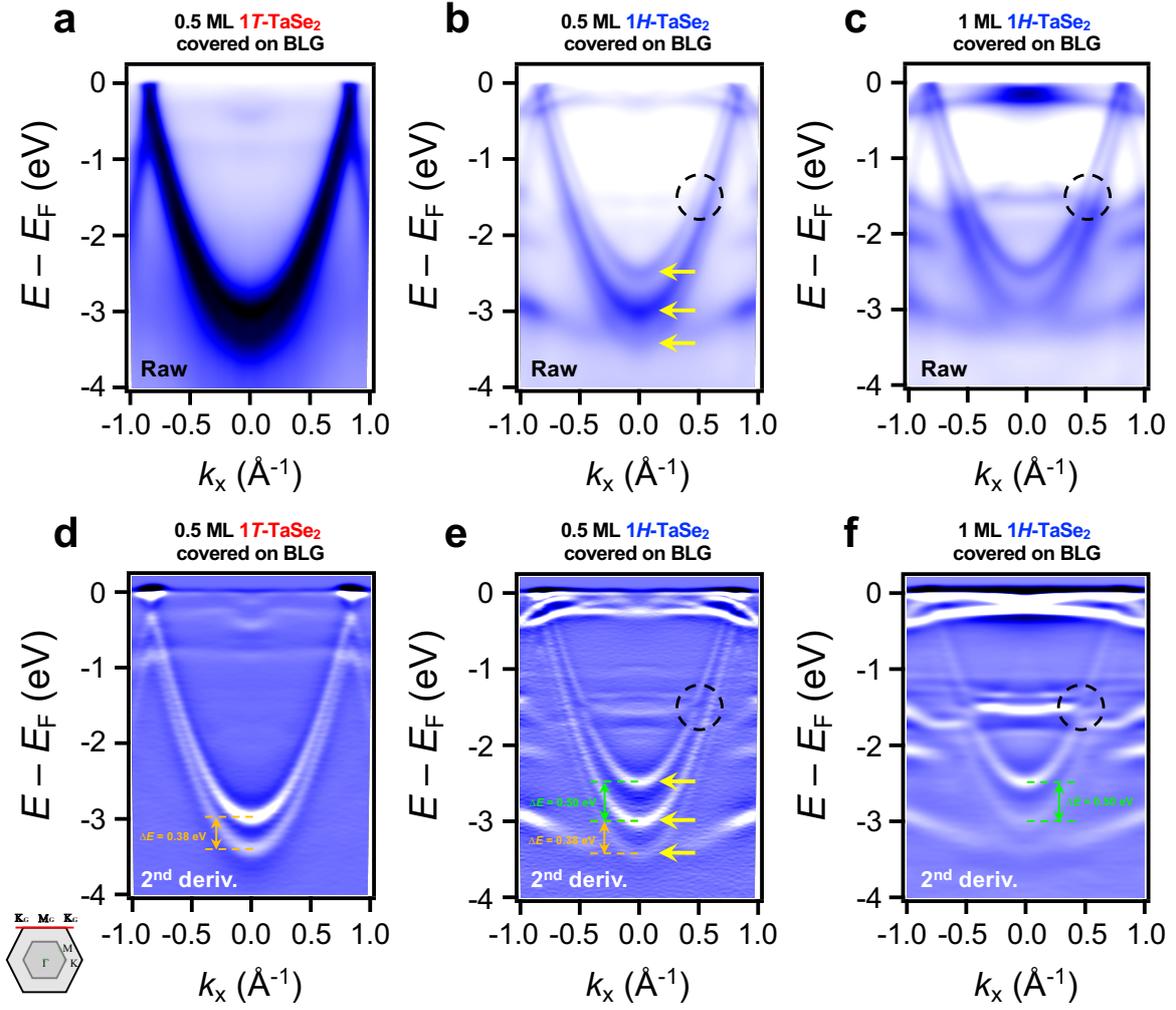

**Figure 3 Comparison of the effect of the crystal structure of ML TaSe$_2$ films on the BLG substrate**. **a-c,** ARPES data of BLG π bands taken along the K$_G$–M$_G$–K$_G$ direction of the BZ of BLG. **d-f,** The second derivatives of ARPES data in panels **a-c**. All ARPES data were taken using *p*-polarized photons at 10 K to better visualize the BLG π bands. The black dashed circles denote the kinked structures of BLG π bands. Yellow arrows represent the split of BLG π bands at the M$_G$ point. Green and orange arrows, and dashed lines indicate the splitting size of the BLG π bands (ΔE).

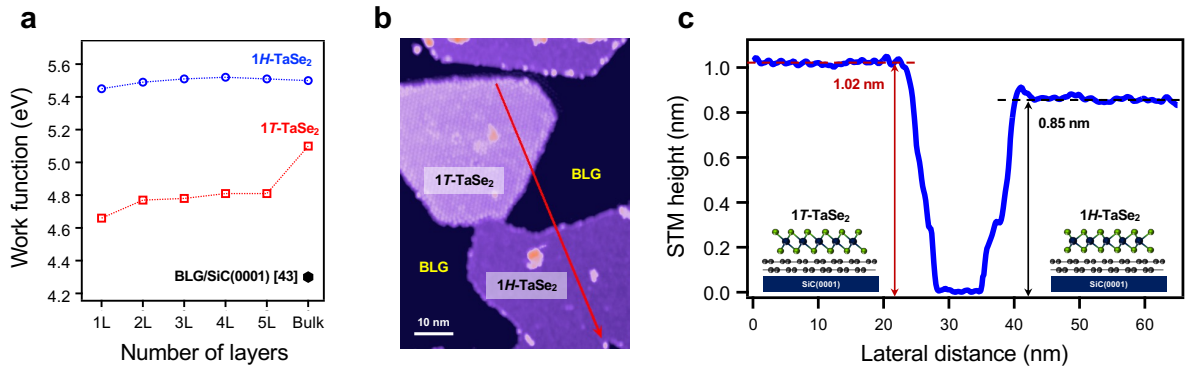

**Figure 4 Thickness-dependent work function and STM step height of ML TaSe$_2$ films on BLG**. **a**, The calculated work function of few-layer 1$T$-TaSe$_2$ (red) and 1$H$-TaSe$_2$ (blue). **b**, STM topographic image with islands of both ML 1$T$-TaSe$_2$ (light purple) and 1$H$-TaSe$_2$ (deep purple) on a BLG/SiC(0001) substrate (scanned at sample bias $V_b$ = -1 V and tunnelling current $I_t$ = 5 pA at 5 K). **c**, An STM height profile taken along a red arrow shown in panel **b**.